\documentclass[12pt,preprint]{aastex}
\usepackage{natbib}
\usepackage{graphicx}
\usepackage{multirow}
\usepackage[usenames]{color}
\usepackage{amssymb}

\newcommand{\hi}{\mbox{H\,\small I}}
\newcommand{\kms}{\mbox{km\ s}$^{-1}$}
\newcommand{\W}{W_{50}}

\newcommand{\subhi}{_{\rm{\scriptstyle{H}\,\scriptscriptstyle I}}}
\newcommand{\Mhi}{M\subhi}
\newcommand{\Fhi}{F\subhi}
\newcommand{\Flim}{F_{\rm{\scriptstyle{H}\,\scriptscriptstyle I,lim}}}

\newcommand{\alf}{\mbox{ALFALFA}}
\newcommand{\msun}{M_\odot}
\newcommand{\ici}{C_{59}}
\newcommand{\rici}{C_{59,r}}
\newcommand{\diam}{D_{25}}
\newcommand{\rdiam}{D_{25,r}}
\newcommand{\rba}{(b/a)_{r}}
\newcommand{\univ}{\hat{\mbox{\boldmath $e$}}}

\slugcomment{Accepted for publication in The Astrophysical Journal}
\shorttitle{\hi\ CONTENT AND OPTICAL PROPERTIES OF \alf\ GALAXIES. II.}
\shortauthors{TORIBIO ET AL.}

\begin{document}

\title{\hi\ CONTENT AND OPTICAL PROPERTIES OF FIELD GALAXIES FROM THE
  \alf\ SURVEY. II.\ MULTIVARIATE ANALYSIS OF A GALAXY SAMPLE IN
  LOW DENSITY ENVIRONMENTS}

\author{M.\ Carmen Toribio and Jos\'e M.\ Solanes}
\affil{Departament d'Astronomia i Meteorologia and Institut de Ci\`encies del
Cosmos, Universitat de Barcelona, Mart\'{\i} i Franqu\`es 1,
08028~Barcelona, Spain}
\email{mctoribio@am.ub.es, jm.solanes@ub.edu}
\and
\author{Riccardo Giovanelli, Martha P. Haynes and Ann M.\ Martin}
\affil{Center for Radiophysics and Space Research, Space Sciences Building,
Cornell University, Ithaca, NY 14853}
\affil{National Astronomy and  Ionosphere Center,
 Cornell University, Ithaca, NY 14853 }
\email{riccardo@astro.cornell.edu, haynes@astro.cornell.edu, amartin@astro.cornell.edu}

\begin{abstract}

  This is the second paper of two reporting results from a study of the 
  \hi\ content and stellar properties of nearby galaxies detected by the
  Arecibo Legacy Fast ALFA blind 21-cm line survey and the Sloan Digital  
  Sky Survey in a 2160 square degree region of high galactic latitude sky 
  covered by both surveys, in the general Virgo direction. Here we analyze 
  a complete \hi\ flux-limited subset of 1624 objects with homogeneously 
  measured 21-cm and multi-wavelength optical attributes extracted from the
  control sample of \hi\ emitters in environments of low local galactic 
  density assembled by \citet{Tor10}. Strategies of multivariate data analysis 
  are applied to this dataset in order to:
  i) investigate the correlation structure of the space defined by an
  extensive set of potentially independent observables describing gas-rich
  systems; ii) identify the intrinsic parameters that best define
  their neutral gas content; and iii) explore the scaling relations
  arising from the joint distributions of the quantities most strongly
  correlated with the \hi\ mass. The principal component analysis
  performed over a set of five galaxy properties reveals that they are 
  strongly interrelated, supporting previous claims that nearby \hi\ emitters 
  show a high degree of correlation. The best predictors for the expected
  value of $\Mhi$ are the diameter of the stellar disk, $\rdiam$,
  followed by the total luminosity (both in the $r$-band), and the
  maximum rotation speed, while morphological proxies such as color
  show only a moderately strong correlation with the gaseous content
  attenuated by observational error. Among the various inferred prescriptions, 
  the simples and most accurate is 
  $\log(\Mhi/\msun)= 8.72 + 1.25\log(\rdiam/\mbox{kpc})$. 
  We find a slope of $-8.2\pm 0.5$ for the relation between optical magnitude 
  and log rotation speed, in good agreement with Tully-Fisher studies, as well 
  as a log slope of $1.55\pm 0.06$ for the \hi\ mass-optical galaxy size relation. 
  Given the homogeneity of the measurements and the completeness of our dataset,
  the latter outcome suggests that the constancy of the average (hybrid) \hi\ 
  surface density advocated by some authors for the spiral population is just 
  a crude approximation.

\end{abstract}

\keywords{ PACS: 02.50.Sk, 98.52.Nr, 98.62.Ai, 98.62.Lv, 98.62.Qz,
  98.62.Ve}
\maketitle

\section{INTRODUCTION}

While the literature abounds with attempts of improving our knowledge
about the formation and evolution of galaxies from the
cross-correlation of the main properties of their baryonic components
\citetext{e.g., \citealt*{GPB96,RSP-B05,Gar-A09}, to name a few
  representative examples}, the possibility of using this sort of
relationships to set reference standards for the \hi\ content has
received comparatively less attention. Early comparisons of the
neutral hydrogen abundance between Virgo cluster and field galaxies
put the emphasis on using distance-independent measures, such as the
$\Mhi/L$ and $\Mhi/D^2$ ratios \citep[e.g.,][]{DL73,CBG80}, $L$ and
$D$ being, respectively, the optical luminosity and intrinsic linear
diameter at a certain wavelength. \citeauthor*{HG84}~(\citeyear{HG84};
hereafter \citeauthor{HG84}) were the first both to carry out an
objective evaluation of the performance of different diagnostic tools
for the \hi\ content and to provide a rigorous operational definition
of this quantity. With the help of a control sample of 288 galaxies
with 21-cm line emission belonging to the Catalogue of Isolated
Galaxies \citep[CIG,][]{Kar73}, these authors demonstrated that,
whatever the Hubble type, the optical linear diameter is the most
important diagnostic tool for the \hi\ mass of galaxies. New
expressions for the standards of \hi\ content according to
\citeauthor{HG84}'s definition were later derived in an unbiased way
by \citet*{SGH96} from a larger, integrated \hi\ flux-limited sample
of 532 galaxies from the Catalog of Galaxies and Clusters of Galaxies
\citep[CGCG,][]{Zwi6168} located in the lowest density environments of
the Pisces-Perseus supercluster region.

One serious limitation of these studies, carried out in a time when
wide-field redshift surveys were still in their infancy, is that they
had to deal with heterogeneous datasets of optically selected targets
assembled from incomplete catalogs and affected by complex sampling
biases that undermined the validity of the results. In this paper and
the accompanying one \citetext{\citealt*{Tor10}; hereafter Paper I},
we conduct a systematic analysis of the main structural properties of
galaxies selected according to their \hi-line emission, which in terms
of both sampling quality and statistics represents a significant
improvement with respect to earlier studies of this kind. Our study
grows out from the combination of data from two large surveys that
homogeneously map the distribution of extragalactic sources over a
significant fraction of the local universe. These are a compilation of
all the data from the ongoing Arecibo Legacy Fast ALFA Survey (\alf)
blind 21-cm line survey \citep{Gio05} gathered so far in the northern
Galactic hemisphere, which contains \hi\ measurements distributed in
two separate regions of the high Galactic latitude sky that cover a
$z$-space volume of about 2160 deg$^2$ $\times$ $18,000$ \kms, and the
Sloan Digital Sky Survey Data Release Seventh \citep[SDSS
DR7;][]{Aba09}, which is complemented with additional data from the
NYU Value-Added Galaxy Catalog \citep[NYU-VAGC;][]{Bla05}.

In Paper I, we deal with the assembly of control samples of \alf\
galaxies that are expected to show little or no evidence of
interaction with their surroundings and, therefore, that are suitable
for providing absolute measures of the \hi\ mass. According to the
results of this study, the optimal dataset to set up standards for the
neutral gas content of galaxies is a sample of 5647 \hi\ emitters
found in regions of low local galactic density, as defined by a
nearest neighbor approach. A complete 21-cm flux-limited subset of
this control sample will be used here (Section~\ref{data_selection})
with the aim of exploring inter-variable linear
correlations\footnote{The correlation analysis carried out in this
  work requires that we ignore any possible curvature in the
  relationships investigated.} and determining the combinations of
intrinsic properties that best define the \hi\ mass of gas-rich
objects. In the remainder of the present manuscript, we will show
first that the galactic stellar size, luminosity, rotation speed, and
to a lesser extent the color, are the intrinsic factors most closely
related to the \hi\ mass (Section~\ref{parameter_selection}), Then, we
will apply strategies of non-parametric multivariate data analysis to
these variables in order to: determine their correlation structure and
the number and orientation of the statistically significant principal
components (Section~\ref{PCA}); establish standards of normalcy for
the \hi\ content of galaxies (Section~\ref{HI_standards}); and examine
the constraints that can be inferred from these relationships on the
most firmly established empirical scaling laws for disk galaxies
(Section~\ref{planar_correlations}). Appendix~\ref{completeness},
briefly discusses two statistical tests implemented in order to assess
the completeness limits of the \alf\ data as a function of integrated
\hi\ flux.

The luminosity distances needed to calculate the various
distance-dependent structural properties used in this investigation
have been inferred within the framework of the standard concordant
flat $\Lambda$CDM cosmology with a reduced Hubble constant
$h=H_0/(100$\,\kms$\,$Mpc$^{-1})=0.7$.

\section{SAMPLE SELECTION}\label{data_selection}

We use a trimmed version of the Low-Density Environment (LDE) \hi\
galaxy sample assembled in Paper I. The original LDE sample consists
of 5647 reliable \alf\ detections with a signal-to-noise ratio
$S/N>4.5$ that have an optical counterpart in the SDSS catalog and
that inhabit regions of low local galactic density ($\rho_6\leq
0.5\,h^3$ galaxies~Mpc$^{-3}$; see Paper I for details). In order to
deal with data of the highest completeness and quality, the present
analysis has been restricted, however, to those \hi\ sources
designated code 1, i.e., with a $S/N > 6.5$, a clean spectral profile,
and a good match between the two polarizations independently observed
by \alf, for which it can be assumed that the completeness limit
is well represented by the detection limit of the survey. Tests of the
performance of the signal detection pipeline made by \citet{Sai07}
estimate a reliability close $100\%$ for \alf\ objects above the
prescribed $S/N$ threshold and an overall completeness approaching
$\sim 90\%$ for those with a narrow observed velocity width ($< 150$
\kms). In contrast, for the few code 2 sources originally included in
the LDE dataset, which have $4<S/N<6.5$, the reliability for
detections of narrow signal is reduced to values near $60\%$ and the
overall completeness to $\sim 70\%$.

Likewise, we have removed from the parent LDE sample those galaxies
located beyond $15,000$ \kms, where the \alf\ survey detection ability
drops significantly due to a strong radio frequency interference (RFI)
signal from the San Juan airport FAA radar. (We remind the reader that
the original LDE dataset is already confined to objects beyond 2000
\kms\ in order to eliminate the sources with the most uncertain
distances and, in particular, a great deal of the kinematic influence
of the Virgo cluster.)  

The large size of the LDE sample has allowed us to be also demanding
when selecting galaxies with favourable inclination angles. Thus, we
have taken into account that at low apparent inclinations orientation
estimates become more uncertain and are skewed towards larger values
by nonaxysimmetric features in the images, ultimately leading to
deprojected properties with divergent errors as a face-on orientation
is approached, an issue well known in Tully-Fisher studies
\citep[e.g.][]{TS09}. Furthermore, the extinction and reddening
corrections that must be applied to optical observables are stronger
and less reliable for nearly edge-on objects. Accordingly, we have
chosen to restrict the present analysis to the 3032 LDE galaxies that,
in addition to verifying the above conditions, have also an isophotal
$r$-band axial ratio $0.85\geq (b/a)_r \geq 0.2$ (equivalent to
$30\degr \leq i \leq 80\degr$ for disks of negligible thickness). A
detailed analysis of the inclination dependence of the different
relationships examined in this work corroborates that all the
correlation coefficients vary little within this range of
$b/a$. Furthermore, we have attempted to reduce as much as possible
the impact of possible outliers on the inferred relationships, which
could be significant especially for low \hi-mass objects, by excluding
galaxies with highly inaccurate SDSS measurements. Although the
restrictions on the axial ratio already discard most of the potential 
optical outliers, an additional $\sim 7$ per cent of the remaining
sources were eliminated for this reason.

Last but not least is the fact that \alf\ is a noise-limited survey,
with a sensitivity that depends on the observed source's \hi\
linewidth, while the correlations that we want to study ideally ought
to be inferred from a volume-limited sample. To sidestep the natural
bias of blind 21-cm surveys against sources with low fluxes and large
velocity widths, we will be dealing with a subset of the LDE sample
that includes objects brighter than a stringent integrated \hi\ flux
$\ge 1.3$~Jy \kms, for which the survey can be considered complete in
a statistical sense regardless of line width (see
Appendix~\ref{completeness}). Each galaxy in this restricted dataset
will be weighted by the inverse of the maximum \emph{effective}
volume, $V^\prime_{\rm max}$, in which it should, on average, have
been detected. To calculate the latter, we have taken into account not
just the observed integrated flux of the source, its distance, and the
adopted sensitivity limit, but also the presence of large-scale
structure in the surveyed volume and the loss of signal that result
from man-made RFI, which alter locally the survey completeness and
therefore the detection probability of the sources. In the remaining
of the paper, we will refer to this selection as our \hi\ flux-limited
LDE 'high-quality' galaxy sample (LDE-HQ for short), which totals 1624
gas-rich objects.

\section{PARAMETER SELECTION}\label{parameter_selection}

Central to this work is the search of dependencies between the neutral
gas mass content and other intrinsic properties of \hi\ emitters. We
have selected the largest possible number of observational parameters
that, besides being suitable to characterize gas-rich objects, are
inferred from good quality measurements and either publicly available
or easy to compute.

\subsection{Available Parameters}\label{all_params}

In our quest for the most relevant galaxian properties that define the
\hi\ content, we first review the extensive set of radio and optical
parameters that can be inferred from the observables listed by the
\alf\ and SDSS DR7 and NYU-VAGC catalogs, which in some cases provide
different estimates of the same attribute (e.g., Petrosian and model
magnitudes to measure brightness, isophotal diameter and Petrosian
radius to measure the angular size, and so on). After a first
screening of all the possible variables available, paying attention to
factors such as the relative size of the obsevational errors, as well
as the suitability and robustness of the measurements for the kind of
galaxies under scrutiny (i.e., blue, late-type, star-forming systems),
we select the following properties as the most convenient for our
study.

\begin{itemize}
\item The two main measures that can be obtained from \alf\
  observations: the 21-cm linewidth of the source at the $50\%$ level
  of the two peaks, $\W$, and the \hi\ mass, which is estimated from
  the equation
  \begin{equation}\label{himass}
    \Mhi=2.356\times 10^5 d^2 \Fhi\;, 
  \end{equation}
  where $\Fhi$ is the 21-cm line flux integral expressed in Jy \kms\
  and $d$ is the cosmological distance to the source in Mpc calculated
  from the multiattractor flow model of local peculiar velocities
  developed by \citet{Mas05}. Examination of the variation of \hi\
  surface density with axial ratio for \alf\ galaxies has lead us to
  neglect the effects of internal \hi\ self-absortion on $\Fhi$. Note
  that in this paper $\W$ represents the \emph{intrinsic} width
  corrected not only for inclination, but also for the effects of
  redshift broadening and turbulence \citep[see][]{Spr05}.

  Due to the absence of reliable estimates of the intrinsic axial
  ratio $q$ of the targets, we choose to use the $r$-band isophotal
  axial ratio $\rba$ from the SDSS as a proxy for inclination in the
  calculation of the intrinsic linewidths. The error on inclination
  corrections arising from taking $q=0$ is in general negligible,
  except for nearly edge-on objects, which have been removed from
  our dataset (see Section~\ref{data_selection}).
\item The luminosities (and their corresponding absolute magnitudes)
  in the five SDSS bands from Petrosian apparent magnitudes. The
  latter, which are especially suited for bright, extended objects,
  lead to recover almost all the light from late-type galaxies and
  around $80\%$ for early types \citep{Bla01}. Absolute magnitudes
  have been corrected to face-on values following \citet{Sha07}. They
  are not corrected for the seeing effect.
\item The 25 mag~arcsec$^{-2}$ isophotal major-axis diameter, $\diam$,
  in the five SDSS bands, which provide a more continuous measure of
  the scale of galaxies than the Petrosian radii $R_{50}$ and
  $R_{90}$. The minimum velocity cutoff of 2000 \kms\ adopted in Paper
  I when defining the LDE sample leaves out of the present analysis
  more than half of the mostly nearby, blue faint objects having
  unrealistically small isophotal angular radii in the SDSS database.

  Isophotal linear diameters have been corrected for inclination by
  using transformations of the form
  \begin{equation}\label{Diam}
  \log \diam=\log \diam^{\rm{obs}} + \beta \log (b/a)\;,
  \end{equation}
  where $\diam$ and $\diam^{\rm{obs}}$ are, respectively, the
  intrinsic and observed values of this variable in a given band,
  whereas the coefficient $\beta$ measures the strength of the
  corresponding attenuation. We note that our corrections are somewhat
  stronger than those estimated by \citet{Mal09} from
  equation~\ref{Diam} for $R_{50}$. For instance, we obtain
  $\beta_{r}=0.35$ in the $r$-band, while the attenuation for
  $R_{50,r}$ calculated by the latter authors is 0.20.
\item The colors from model magnitudes. Here, we explore the
  combinations $(u-g)$, $(g-r)$, $(r-i)$, $(i-z)$, $(u-r)$, $(u-i)$,
  $(u-z)$, $(g-i)$, $(g-z)$, and $(r-z)$. 

  The effect of inclination on the colors of our galaxies has been
  corrected by using the variation of color with axial ratio quoted in
  \citet{Mas10}. For galaxies with $\log(a/b)<0.7$, we use the
  straight-line fits listed in their Equation (3), inferred for Galaxy
  Zoo (GZ) spirals with $R_{90}>10\arcsec$ and a constant extension
  of these fits for objects with $\log(a/b)>0.7$, as measured in
  $g$-band. We have not applied the strongest color corrections
  derived by \citeauthor{Mas10} for systems with smaller apparent
  sizes, because we have verified that the adopted transformation is
  enough to make the average intrinsic $(g-r)$ color of \alf\ sources
  independent of viewing angle.
\item The S\'ersic index, $n$, from the NYU-VAGC catalog, which
  measures the shape of the observed $r$-band luminosity profile of a
  galaxy fitted using the S\'ersic $R^{1/n}$ formula with elliptical
  isophotes. Available only for galaxies with $r < 18$~mag.
\item The (inverse) index of light concentration, $\ici =
  R_{50}/R_{90}$, in the five SDSS bands, which ranges from 0 to 1 and
  is available for the full SDSS DR7 dataset. Not corrected for
  seeing.
\end{itemize}

\begin{figure}
\begin{center}
\epsscale{0.8}
\plotone{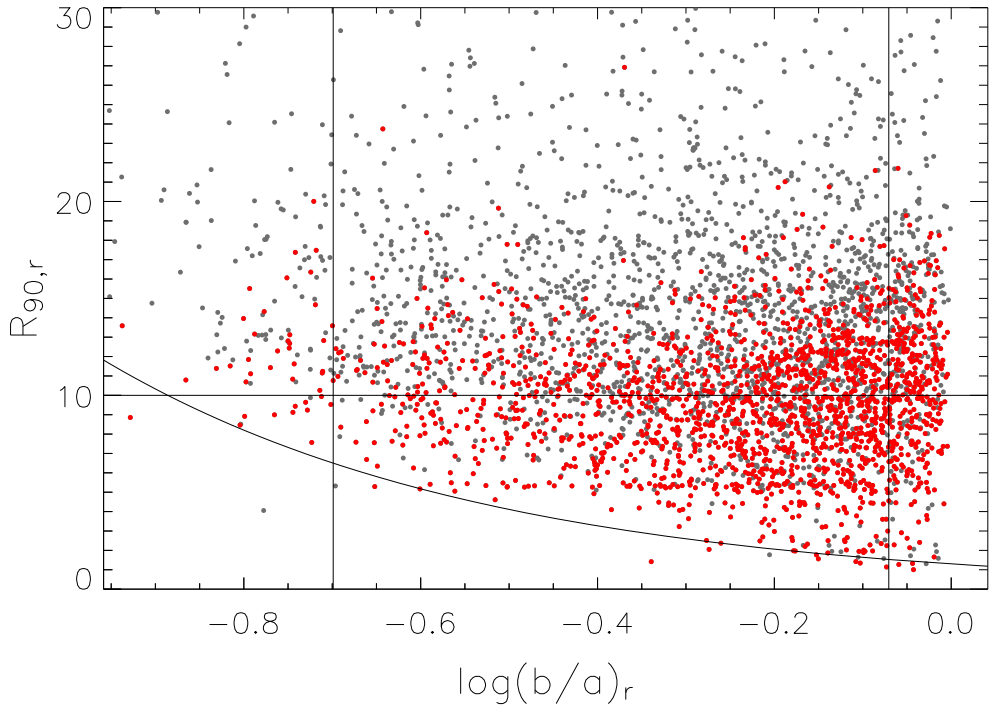}

\end{center}
\caption{The $r$-band axial ratio versus the radius enclosing 90 per
  cent of the $r$-band light for galaxies in the \alf\ LDE
  sample. Vertical lines correspond to the limits in axial ratio
  adopted in the definition of the LDE-HQ sample (in red, objects with
  $\Fhi < 1.3$ Jy~\kms, which we also exclude from the correlation
  analysis). The horizontal line, plotted at $R_{90,r} = 10\arcsec$,
  marks the limiting radius above which the seeing effects disappear
  for optical data, while the curved line shows a model in which the
  minimum possible semiminor axis is $b=1.3\arcsec$, an estimate of
  the 25th percentile best seeing in the SDSS \citetext{see
    \citealt{Mas10}}.}\label{fig_seeing}
\end{figure}

Along with these variables, we also include the isophotal $\rba$,
which given its apparent nature should be uncorrelated with any of the
former quantities and can act therefore as a control parameter. The
values of this axial ratio quoted in the SDSS are not corrected for
seeing, which makes the most inclined galaxies appear rounder than
they should be. We note, however, that the possible impact of seeing
in our data is likely to be insignificant since its effects are only
noticeable for small, highly-inclined galaxies, which are mostly
excluded from our sample due the adopted integrated flux cut (the
members of the LDE-HQ dataset are regular gas-rich galaxies with
$\Mhi\gtrsim 10^{9}\,\msun$). Besides, blind \hi\ surveys in which the
detection probability depends on the observed linewidth like \alf\ are
naturally biased against this kind of targets \citetext{compare, for
  instance, the data point densities in our Figure~\ref{fig_seeing}
  and those in Figure 1 from \citealt{Mas10} derived for GZ spirals}.

Note also that in the present analysis, the color, as well as the
S\'ersic and light concentration indexes, are used as objective
proxies of morphology. Attempts to work with 'classical' indicators of
morphological type, such as the continuous de Vaucouleurs numerical
code listed in the HyperLeda database, have been thwarted by lack of
completeness: about half of the SDSS galaxies do not have information
on their Hubble types, while the same is true for $\sim 16\%$ 
of the \alf\ sources.

In order to examine the correlation structure, we feed in the
logarithms of all these basic variables but $\ici$ and $\rba$. In this
manner, we should be able to find any scaling law that might exist
among them (this is also a must when variables have a lognormal
distribution). This means, in particular, that it is not necessary to
explicitly include in the analysis interesting composite parameters,
such as the stellar mass given by \citet{Gav08}, $\log
(M_\ast/M_\odot)=-0.152+0.518\,(g-i)+\log L_i$ \citetext{a similar
  formula based on the $(g-r)$ color is provided by \citealt{Bel03}},
that are linear combinations of (the logarithm of) two or more single
input variables. This is also the case for the mean surface
brightness, isophotal or Petrosian, $\bar{\mu}\,(\equiv M + 5\log R)$,
which can likewise be expressed as a function of two of the
measurements listed above.

\subsection{Relevant Parameters}\label{best_params}

Before we proceed with the present study there is, however, an
important consideration that must be taken into account. It has to do
with the fact that on a t-test of significance the large size of our
sample results in very low critical values for the Pearson's
correlation coefficients, $r_{\rm P}$, i.e., for the elements of the
various correlation matrices that are inferred in this section. For
instance, $r_{\rm P}^{\rm crit} \sim 0.06$ for a level of significance
of 0.01 on a two-tailed test with about 1600 degrees-of-freedom. This
implies that variables that \emph{ideally} should be essentially
independent of each other might end up exhibiting a very weak, but
nonetheless statistically significant, linear relationship according
to this test, even in the presence of attenuation from measurement
errors. For this reason, as well as to avoid working with an excessive
number of parameters, we have decided to consider that two given
properties are truly connected in practice only if they have a
Pearson's $r$ that, aside of being statistically significant, is also
minimally strong ($\left|r_{\rm P}\right| > 0.3$).

The correlation matrices inferred from different subsets of the
attributes selected in Section~\ref{all_params} demonstrate that
colors and luminosities (absolute magnitudes) are strongly correlated
among themselves, as was to be expected. This allows us to discard all
but one of the colors and all but one of the luminosities. Among these
photometric variables, the ones showing the largest correlation
coefficients with the \hi\ mass are the $g$, $r$, and $i$ band
luminosities, as well as all the optical colors that can be obtained
from combinations of them. Given that the photometric errors in these
three bands are rather similar \citep{Str01}, we have taken into
account both the dynamic ranges of the different colors and the
economy in the number of involved bands to finally select the $r$-band
luminosity and the $(g-r)$ color as the most adequate representatives
of these two fundamental stellar properties.

The correlation analysis has also evidenced that the available
measurements of the isophotal diameter and the S\'ersic index in the
five SDSS bands are degenerated. As before, the superior quality of
the SDSS photometry in the central $g$, $r$, and $i$ bands results in
somewhat stronger correlations of these variables with \hi\ mass at
such wavelengths. Consistency with the adopted bandpass for measuring
the amount of light, has lead us to select $r$-band estimates of
$\diam$ and $n$ to represent these quantities too.

Overall, we find that the isophotal $r$-band diameter, the $r$-band
luminosity, and the $\W$ linewidth are tightly related to the
$\Mhi$. A second group of attributes, the $(g-r)$ color and the
$r$-band S\'ersic index, are moderately aligned with this quantity
\citetext{$\left|r_{\rm P}\right|\sim 0.3$--0.5; see \citealt{Dis08}
  for a similar conclusion regarding color}. While we have decided to
retain $(g-r)$ in the set of parameters that may be needed to describe
the cold gas content of a galaxy, we have chosen to omit from this
list the S\'ersic index, as it has a somewhat smaller $\left|r_{\rm
    P}\right|$ and is available only to galaxies in the NYU-VAGC
catalog (i.e., with $r\leq 18$~mag). Finally, the third morphological
separator selected, the index of light concentration, shows little or
no indication of a major dependence on the \hi\ mass, exhibiting a
Pearson's $r$ of similar strength ($\lesssim 0.2$), for instance, that
correlations involving the apparent inclination, so it will also be
discarded in the remainder of this study.

\section{RELATIONSHIPS AMONG \hi\ AND OPTICAL PROPERTIES}\label{correlation_structure}

We now complete the characterization of gas-rich galaxies by
circumscribing ourselves to the final set of five non-degenerate,
intrinsic properties selected in the previous section.

To begin with, we will examine in detail the correlation structure of
this multivariate space to both identify its most statistically
significant principal components and their degree of alignment with
the selected observed parameters. We will then use the correlations
among the measured variables to derive formulae for predicting the
neutral hydrogen content according to two different approaches. On the
one hand, we will seek for the most probable value of $\Mhi$ assuming
the other four properties are precisely known. This involves solving a
multivariate regression problem in order to obtain equations useful to
establish standards of normalcy for the \hi\ content of galaxies ---we
shall define as 'normal' the typical values of our LDE-HQ galaxy
sample members--- from a set of diagnostic parameters easily
accessible to observation. On the other hand, we will also determine
the best-fitting axes of all the different pairs that can be built
from these five basic attributes and the constraints they impose on
the scaling laws connecting fundamental properties of galaxies.

\subsection{PCA Results}\label{PCA}

Here we deal with the correlation matrix of the variables $\log
(\Mhi\,[\msun])$, $\log (\W\,[$\kms$])$, $\log
(\rdiam\,[\mbox{kpc}])$, $M_r\,[\mbox{mag}]$, and
$(g-r)\,[\mbox{mag}]$ in order to perform a principal component
analysis (PCA) in this parameter space. Unlike the covariant matrix,
the use of the correlation matrix entails the standardization of the
original variables putting them on an equal footing: all samples of
variables are get to have zero mean and unit standard deviation. This
scaling of the measurements avoids the creation of spurious
interrelations arising from the preponderance (i.e., larger dynamic
range) of certain properties. For the PCA calculations, we employ the
IDL procedure
\emph{pca.pro}\footnote{\texttt{http://idlastro.gsfc.nasa.gov/ftp/pro/math/pca.pro}.},
implemented following the description by \citet{MH87}. This algorithm
has been modified in order to account for the flux limitation of our
sample and the measurement error of the estimates.

As stated in Section~\ref{data_selection} (see also
Appendix~\ref{completeness}), the integrated flux cutoff of the LDE-HQ
sample has been compensated by weighting each of its member galaxies
by the inverse of the maximum volume, $V^\prime_{\rm max}$, in which
it could have been observed corrected for the systematic effects of
large-scale structure and loss of signal due to RFI. This correction
is larger for the lowest \hi-mass objects since they are only detected
at short distance\footnote{The LDE-HQ dataset represents a
  volume-limited sample of more than $29,000$ galaxies.}. We have also
taken into account that the correlations between variables are
weakened in the presence of measurement error. Given two sets of
estimates $\hat{X}$ and $\hat{Y}$ \emph{with independent measurement
  errors}, the disattenuated correlation between the underlying
variables $X$ and $Y$ can be obtained from the formula \citetext{cf.\
  \citealt{Spe04}}
\begin{equation}\label{eq_spearman}
r_{\rm P}(X,Y)=\frac{r_{\rm P}(\hat{X},\hat{Y})}{\sqrt{R_{XX}R_{YY}}}\;,
\end{equation} 
where the reliability coefficients $R_{XX}$ and $R_{YY}$ are defined
as one minus the ratio between the variance on the corresponding
measurement error and the total observed variance. In practice, this
means that variables are standardized using an estimate of their true
standard deviation, instead of the observed one. We have followed
\citet{FH78} \citetext{see also \citealt{BP75}} and extended this
correction to the multivariate case by imposing the constraint of
producing a valid correlation matrix for our disattenuated variables,
i.e., a matrix that is at least positive-semidefinite, after checking
that their associated measured error scores are essentially
independent.

The results of our PCA analysis are summarized in
Tables~\ref{tab_pca_att} and \ref{tab_pca} in the form of the
$1/V^\prime_{\rm max}$-weighted correlation matrix, all its
eigenvectors and eigenvalues, i.e., the variances of the data in the
directions of the principal axes, as well as the rms residuals between
the 5-dimensional space (manifold) of the observations and the
different $p$-dimensional subspaces ($p=1,...,5$) that best describe
them. Table~\ref{tab_pca_att}, which presents the results for
disattenuated correlations, also lists the adopted estimate of the
typical measurement error and the square root of the reliability
coefficient for the observables. In general, the selected parameters
have a near perfect reliability, except the color, for which the
attenuation correction is significant.

The correlation matrices indicate the existence of large linear
correlations ($|r_{\rm P}|> 0.60$) for all pairs of variables except
for the \hi\ mass and color, that exhibit correlation coefficients of
medium size ($|r_{\rm P}|\sim 0.45$--0.60; note, in contrast, the
substantially higher coefficients of the correlations between
luminosity and color). Table~\ref{tab_pca_att} shows that the five
galactic attributes selected are all well correlated with the first
principal component. The latter is endowed with direction cosines of
nearly equal absolute value ($\approx 1/\sqrt{5}\approx 0.45$) and has
an associated eigenvalue $\lambda_1$ of $\sim 4.2$, out of a maximum
possible of 5.0, implying that about $83\%$ of the total variance in
the adopted 5-parameter space can be explained by a single principal
axis. The second, and already statistically insignificant, principal
component, draws mostly from $\Mhi$, and accounts for an additional
$\sim 10\%$ of the global variance, while a linear subspace of three
dimensions is enough to explain the practical totality ($\sim 99\%$)
of it (these numbers are reduced somewhat when we do not use
disattenuated correlation estimates; see Table~\ref{tab_pca}). The
principal 4-plane brings the rms residuals down to the level of the
adopted observational errors with the exception of $M_r$, whose
variance cannot be fully accounted for. This might indicate either
that there are still hidden parameters controlling the structure of
disk galaxies, such as perhaps the mass-normalized star formation
rate, which is tightly related to the light concentration index
\citep[e.g.][]{NCA04}, or that the observational error adopted for
this variable has been too optimistic. Note also that the correlation
matrix reveals a very strong negative relationship between the
$r$-band magnitude and isophotal diameter, which is responsible for
the nearly null variance attached to the last eigenvector.

Our finding that one single principal component ---which does not
appear to be dominated by any of the five major properties
investigated--- explains a great deal of the variance of the observed
manifold suggests that the structure of regular gaseous galaxies
should be determined by very few independent features. This is
consistent with the results of earlier PCA-based studies that have
attempted to elucidate the degree of organization shown by disk
galaxies \citep[e.g.][]{Bro73,BGB81,Con06}, as well as in very good
agreement with the conclusion that \hi\ galaxies lie essentially on a
single fundamental line reached in a recent study by \citet{Dis08}
that, like the present one, combines homogeneous 21-cm data (from the
\hi\ Parkes All Sky Survey; HIPASS) with SDSS optical measures. 

Regarding the possibility, suggested by these latter authors, that
\hi-selected galaxies have colors made up of two components, one
\emph{systematic}, correlated with the other variables and the single
principal component, and a so-called \emph{rogue} component, which
only alignes with itself and therefore could act as a second
significant parameter, we offer a different explanation. Our results
---based on a dataset about 8 times larger, though spanning a smaller
dynamic range---, show that when the PCA is carried on the attenuated
(and $1/V^\prime_{\rm max}$-weighted) correlation matrix, we obtain a
second principal component, PC2, well aligned with
color. Nevertheless, once the measurements are corrected from
observational error the color contribution for PC2 significantly
weakens and is no longer the dominant one (compare the corresponding
eigenvectors in Tables~\ref{tab_pca} and \ref{tab_pca_att},
respectively). This leads us to conclude that the possible
statistically significant second degree of freedom found by
\citet{Dis08} is actually an artifact produced by the substantial
attenuation exerted by the measurement error of the latter observable
on the moderately strong intrinsic \hi\ mass-color correlation (see
also Paper I).

\subsection{Standards of \hi\ Content}\label{HI_standards}

We have just shown that the five observables we are dealing with are
actually interconnected. In this situation, any multiple regression
model which attempts to describe the \hi\ mass in terms of
interrelated predictor variables (multicollinearity) would be
associated with an ill-conditioned correlation matrix, i.e., a matrix
whose inversion is numerically unstable (if there were one or more
exact linear relationships among the variables the matrix would not be
invertible). Thus, in the presence of multicollinearity the impact of
the individual predictors on the response variable tends to be less
precise than if the predictors were uncorrelated with one another. To
remedy this problem, we have adopted a two-stage procedure known as
Principal Component Regression \citep[PCR; e.g.][]{Coo07} that first
carries out a PCA of all the predictor variables, and then uses the
resulting principal components ---which are independent, and hence
associated with a correlation matrix of full rank--- together with the
dependent variable in an ordinary least squares regression
fit. Besides, one can take advantage of the initial PCA transformation
to reduce the dimensionality of the data by keeping only those new
variables most correlated with the \hi\ mass \citep[e.g.][]{Jol82}.

\begin{figure}
\begin{center}
\epsscale{0.85}
\plotone{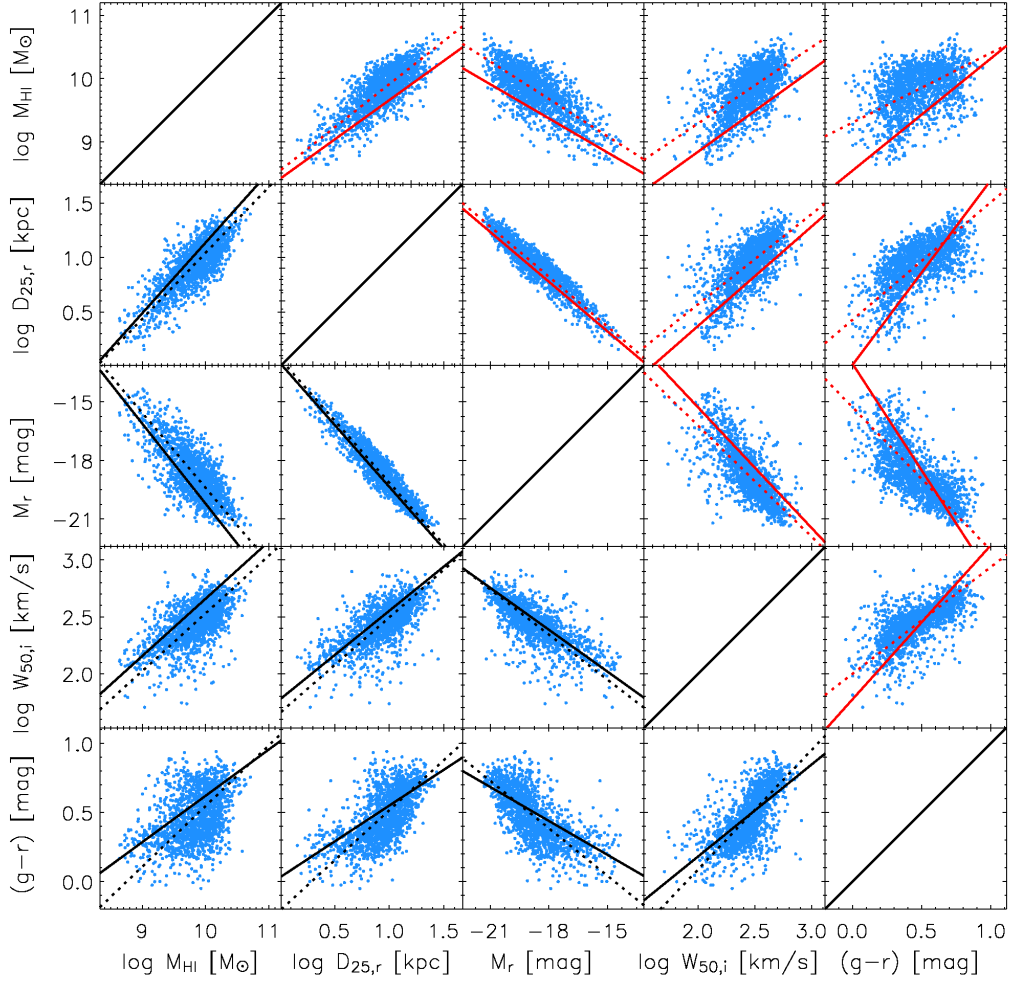}

\vspace{10pt}
\end{center}
\caption{Empirical relations for pairs of properties from LDE-HQ
  data. $1/V^\prime_{\rm max}$-weighted (solid) and unweighted
  (dotted) direct regression fits to the joint distributions are shown
  in red color above the diagonal of the plot, whereas orthogonal fits
  are shown below it. All correlations are corrected for attenuation
  (Equation~\ref{eq_spearman}).}\label{fig_planar}
\end{figure}

We have applied the above procedure to subspaces of increasing
dimension, starting by finding the regression relations between $\Mhi$
and each one of the four remaining properties (see the plots above the
diagonal of Figure~\ref{fig_planar}), and then adding input variables
progressively to seek for the combinations of regressors that best
predict the \hi\ content. This process is stopped when after adding a
new predictor variable the rms residual of the multiple regression
model increases or does not get reduced in an amount comparable or
larger than the typical observational error in $\Mhi$ quoted in
Table~\ref{tab_pca_att}.

In agreement with the results of the previous section, we find that
the best predictions for the \hi\ mass are those depending on a single
regressor variable. Table~\ref{tab_pred} lists, ordered according to
decreasing accuracy as given by the size of the rms residuals, the
central values and associated errors of the coefficients $a_i$ of the
correlations
\begin{equation}\label{eq_pred}
\log\left(\frac{\Mhi}{\msun}\right)= a_0 + a_1\;X 
\end{equation}
for fits with and without $1/V^\prime_{\rm max}$-weighting carried on
disattenuated data. It can be seen that the most precise predictor of
the \hi\ mass is $\rdiam$, just the property most strongly correlated
with it, followed by $M_r$. Among the distance-independent
observables, the best predictor of the \hi\ content is the rotational
width of the disk, whereas the regression model using the $(g-r)$
color is the least accurate, as expected. Of course, the use of the
\hi\ rotation speed to estimate $\Mhi$ only makes sense when the 21-cm
line flux integral is not available and this predictor can be replaced
by proxies like, for instance, the estimators defined in \citet{CHG07}
from optical rotation curves.

The uncertainties of the correlation coefficients quoted in
Table~\ref{tab_pred} include two contributions. The first one is the
random error, calculated by adding in quadrature the statistical error
of the parameter estimates, based on 5000 non-parametric bootstrap
trials of the correlations, and the spread due to observational
errors, which we have calculated through the creation of 10,000
realizations of the correlations after assigning Gaussian random
measurement and distance errors to each galaxy.

The second error term is included to show the impact that voids (and
overdensities of comparable size present in the LDE-HQ sample) can
have on the inferred relationships by systematically undercounting
(overcounting) galaxies in those regions. Remember that we are dealing
with galaxies in low-density environments and that the weighting
scheme depicted in Appendix~\ref{completeness} only corrects for the
redshift-averaged density of galaxies in the surveyed volume. This
systematic error has been evaluated by dividing the sample into 8
equal-area sky regions of $\sim 15\degr$ and calculating the
correlations 8 times, leaving out a different section each time. At
the median survey redshift of $\sim 8000$ \kms, the adopted angular
size corresponds to about 30 Mpc, a scale comparable to the typical
diameter of voids in the local universe \citep{HV04}. The contribution
of galaxy under and overdensities to the variance in any correlation
coefficient $a_i$ is measured through the jackknife error estimator
\citep{Lup93}
\begin{equation}\label{jackknife}
\sigma^2_{a_i}=\frac{N-1}{N}\sum_{j=1}^{N}(a_{i,j}-\overline{a_i})^2\;,
\end{equation}
where $N=8$.

Finally, we have also verified that combinations of two
distance-dependent input variables, which obey equations of the form
\begin{equation}\label{eq_pred_two}
\log\left(\frac{\Mhi}{\msun}\right)= a_0 + a_1\;X_1 + a_2\;X_2\;,
\end{equation}
do not contribute to reduce the spread that may arise from distance
uncertainties. We report in Table~\ref{tab_pred} the coefficients
obtained using the optical size and luminosity as diagnostic
variables, which is also the only multilinear regression model with a
rms residual as good as that of the best linear model (but at the cost
of using two predictors). Looking at the coefficients of this multiple
regression for weighted and unweighted data, it is clear that the
right-hand side of the equations shows a dependence with distance that
neither is null (the ratio $a_1/a_2\neq -5$), as would be expected if
they were defining a surface magnitude, nor does it compensate the
$d^2$-dependence of the \hi\ mass. As done previously for the single
regression models, we account for this breach of distance independence
when calculating the random error of the correlation coefficients.

\subsection{Planar Correlation Diagrams}\label{planar_correlations}

In the previous section, we have been interested in obtaining
predictions for one integral property, the \hi\ mass, from the
observed values of other four ones: size, total magnitude, velocity
width, and color. In this section, we turn our attention to the
constraints that these intrinsic quantities put on the scaling
relations among the fundamental attributes of individual
galaxies. This means that we now consider the five compiled variables
on an equal footing and focus on the correlations arising from the
same PCA technique used in Section~\ref{PCA} applied to pairs of
them. The involved quantities are therefore treated symmetrically,
thus minimizing the inconsistencies that may arise from the possible
'non-commutativity' of the inferred relationships. The coefficients of
the $1/V^\prime_{\rm max}$-weighted and unweighted orthogonal fits of
the form $Y = a_0 + a_1\;X$ between all 10 possible pairs of galaxy
properties are presented in Table~\ref{tab_sca} for disattenuated
data. (For this exercise, the quoted uncertainties depict only the
random error estimates calculated as in Section~\ref{HI_standards}.)
The scatter plots and their best linear fits can be visualized in the
boxes below the diagonal in Figure~\ref{fig_planar}.

The study of the scalings of the most basic properties of galaxies is
central for constraining theories of their formation and evolution. It
has generated an abundant literature, whose detailed revision far
exceeds the scope of the present work. Instead, we have decided to
focus on the comparison between the values for the \emph{mean} slopes
of the strongest correlations we predict, which are those in the
$L(M)RV$ subspace of luminosity (mass), size, and rotation speed, and
those reported in other studies that also combine \hi\ and optical
observations \citetext{\citeauthor{HG84};\ \citealt{SH96}}, or that
specifically study the scalings among the above fundamental properties
in late-type objects \citep{Cou07}. When comparing results allowance
should be made not just for differences in sample size, but also for
other factors such as the waveband of the optical observations, the
specific observables chosen to estimate the above attributes, their
dynamic ranges, or the fitting method employed. Another complication
that distorts the comparison among the different outcomes is the
incompleteness of the datasets that, with the exception of ours, are
all affected by intractable selection biases.

With all these caveats in mind, the agreement between the central
values of the slopes (rounded to the two most significant digits)
reported by the different studies listed above (see
Table~\ref{tab_sca_fun}) can be classified as generally
satisfactory. The largest discrepancies correspond to the correlation
involving the luminosity versus the rotation speed, i.e., the
Tully-Fisher (TF) relation, reflecting the fact that it is always
problematic to accurately find the slope of this empirical law due to
the low dynamic range in $\log V$. The mean values of the log slope
$\gamma$ listed in Table~\ref{tab_sca_fun} range from $\sim 2.6$
\citetext{$-6.5$ in magnitude units; \citeauthor{HG84}} to $\sim 3.4$
\citep{Cou07} and $\sim 3.7$ \citep{SH96} (the first and the last ones
being measured at blue wavelengths and the \citeauthor{Cou07}'s for
the $I$-band). From our weighted data, we get a slope of $3.3\pm 0.2$
(random error) or, equivalently, $-8.2\pm 0.5$ mag, which roughly
falls in the middle of this range and is fully consistent with
estimates reported in TF-specific literature for bright spirals in the
blue/near-IR band \citep[e.g.][]{Wil97,Gio97,Cou97,Mas06}.

Among all the results obtained, the most striking is perhaps the
relationship between the \hi\ mass and the characteristic size of the
stellar distribution of gaseous galaxies, represented here by the
isophotal diameter in the $r$-band, $\rdiam$. Our finding that it has
a central slope $\alpha = 1.55\pm 0.06$ does not support the idea that
\emph{all} \hi-rich galaxies have roughly the same global \hi\ column
density, as recently advocated by \citet[][see also references
therein]{Gar-A09} from a sample of HIPASS galaxies and implied by
correlations such as the one found by \citet{SH96}, listed in
Table~\ref{tab_sca_fun}, from optical size measurements in the
$B$-band. (Implicit in this conclusion is the assumption that \hi\ and
stellar disk sizes are roughly proportional as shown by \citet{BR97}.)
We remind the reader once again that none of these previous studies
based their conclusions on correlation analyses performed on complete
multivariate datasets, as we have done here. In particular, we suspect
that the use of samples largely dominated by late-type spirals, which
are more prone to exhibit a nearly constant mean \hi\ surface density,
may contribute to exacerbate the tendency to find such an
aesthetically appealing result \citetext{see also \citealt{SGH96}}. In
this regard, we wish to emphasize that the claimed constancy of the
mean \hi\ column density by \citet{Gar-A09}, who deal with a sample
that, morphologically speaking, is representative of the entire
population of \hi\ emitters, emanates from a (unweighted) \hi\
mass-$R_{50,g}$ correlation with an actual central slope of $\sim
1.7$, in pretty good agreement with our outcome.

\section{SUMMARY AND CONCLUDING REMARKS}\label{summary}

We have sought for correlations among a large set of extensive 21-cm
and optical homogeneous measures available for the 1624 members of a
complete, \hi\ flux-limited sample of non-clustered, gas-rich
galaxies, not influenced by their environment. Our main aim has been 
identifying the combinations of
intrinsic variables directly arising from observable quantities that
make up the best diagnostic tools for the \hi\ content. The sources
used for this research have been selected from the Low Density
Environment \hi\ galaxy sample of the \alf\ blind \hi\ survey defined
in Paper I. The size of this primary database has been reduced to
include only high-quality \alf\ detections with a $S/N>6.5$ found up
to $15,000$~\kms\ and with moderate inclinations $0.85\geq b/a\geq
0.2$ ($30\degr \leq i \leq 80\degr$ for $q=0$). Furthermore, we have
selected only those \hi\ emitters with an integrated flux $\ge 1.3$~Jy
\kms\ from which the sensitivity of the survey becomes essentially
independent of profile width.

The examination of the correlation structure of these data,
conveniently weighted to compensate for the flux limitation, as well
as for the systematic effects of large-scale structure and loss of
signal due to man-made RFI in the surveyed volume, has produced the
following interesting results.

\begin{itemize}
\item In (bright) gas-rich galaxies the isophotal $r$-band linear
  diameter, total $r$-band luminosity, $\W$ linewidth, and $(g-r)$
  color are the galaxian properties most tightly correlated with the
  total \hi\ mass. The principal component analysis of the manifold
  defined by these variables has revealed relationships with large
  correlation coefficients that are suggestive of a high degree of
  organization in the LDE-HQ sample. This is consistent with the
  idea that \hi\ emitters behave essentially as a one-parameter
  family. As previously noted by \citet{Dis08} \citetext{see also
    \citealt{vdB08} and references therein}, the observed structural
  simplicity of disk galaxies is difficult to reconcile with the
  prevailing theory of hierarchical galaxy formation, which holds that
  the physical properties of these objects are determined by the
  interplay of several potentially independent factors, such as mass,
  spin and (the chaotic at high redshift) merger history of galactic
  halos.

\item In accordance with the output of the PCA, we have found that the
  best predictions for the most probable value of $\Mhi$, assuming
  that the regressor variables are \emph{precisely} known, are those
  depending on a single parameter. Our fits carried on
  $1/V^\prime_{\rm max}$-weighted and disattenuated data show that the
  most accurate predictor for the \hi\ mass of a galaxy is its optical
  diameter, through the equation
  \begin{equation}\label{MHIvsD25}
  \log\left(\frac{\Mhi}{\msun}\right)= 8.72\pm0.06\pm0.06 + 1.25\pm0.06\pm0.07\log\left(\frac{\rdiam}{\mbox{kpc}}\right)\;,
  \end{equation}
  where the first error term is statistical and the second the
  systematic uncertainty arising from the large scale structure
  present in the surveyed volume. In Table~\ref{tab_pred}, we provide
  alternative prescriptions to calculate $\Mhi$ from $M_r$, $\W$, and
  $(g-r)$, as well as from a combination of $\rdiam$ and $M_r$. The
  fact that the models based on the crude morphological indicator
  $(g-r)$ yield rms residuals comparable to the global standard
  deviation associated with $\Mhi$ hints that the morphology of \hi\
  emitters plays a secondary role in determining their neutral gas
  content, as already inferred by \citeauthor{HG84} and \citet{SGH96}.

\item From the joint distributions of the quantities most strongly
  correlated with the \hi\ mass, we derive the mean relationships
  \begin{equation}
  \Mhi\propto R^{1.6}\mbox{,\ }\Mhi\propto L^{0.60}\mbox{,\ }L\propto V^{3.3}\mbox{,\ }R\propto L^{0.40}\mbox{,\ }R\propto V^{1.3}\;,
  \end{equation}
  where $L$ represents the total $r$-band luminosity of the (old)
  stellar disk of a galaxy, and $R$ and $V$ its size and rotation
  speed, respectively. Among the scaling relations that involve \hi\
  measurements, the most interesting ones are the well-known $LV$ or
  TF relation and the ratio between the total neutral gas mass and
  optical radius. For the former, it is noteworthy that we find a
  central slope fully consistent with the typical values reported in
  TF studies at optical/near-IR wavelengths from data that have not
  been specifically selected for this task. On the other hand, the
  slope inferred for the second scaling implies that the hybrid
  surface density of neutral hydrogen ($\propto \Mhi/R^{2}$) is not
  constant, but moderately decreasing with galaxy size. Claims in
  favor of the near universality of the global \hi\ surface density
  for the entire spiral population rely on incomplete datasets biased
  towards galaxies of the latest Hubble subtypes (mostly Sc and Irr),
  for which the constancy of this intensive property is a relatively
  acceptable approximation.
\end{itemize}

To date, most multidimensional statistical studies focusing on the
interrelations among the main properties of galaxies have had to
contend with largely incomplete, heterogeneous samples of modest size
and affected by important selection artifacts. It is therefore evident
that disregarding any of these factors when they are in fact present
may result in inconsistent estimation. In this respect, efforts like
the present one based on the cross-correlation of large datasets
assembled from objective, wide-area surveys with controlled sampling
biases should mark the way forward.

\begin{acknowledgements}
\small

This work was supported by the Direcci\'on General de Investigaci\'on
Cient\'{\i}fica y T\'ecnica, under contract AYA2007-60366. M.C.T.\
acknowledges support from a fellowship of the Spanish Ministerio de
Educaci\'on y Ciencia. RG, MPH, and AMM receive support from NSF grant
AST-0607007 and from a grant from the Brinson Foundation.

We thank the many members of the \alf\ team who have contributed to
the acquisition and processing of the \alf\ dataset over the last six
years.

This research is based mainly on observations collected at Arecibo
Observatory. The Arecibo Observatory is part of the National Astronomy
and Ionosphere Center, which is operated by Cornell University under a
cooperative agreement with the National Science Foundation.

We have also made use of the HyperLeda database
(\texttt{http://leda.univ-lyon1.fr}) and the NASA/IPAC Extragalactic
Database, which is operated by the Jet Propulsion Laboratory,
California Institute of Technology, under contract from the National
Aeronautics and Space Administration. Likewise, we are grateful to all
the people and institutions that have made possible the NYU
Value-Added Galaxy Catalog
(\texttt{http://sdss.physics.nyu.edu/vagc/}) and the Sloan Digital Sky
Survey (SDSS; \texttt{http://www.sdss.org/}). Funding for the SDSS has
been provided by the Alfred P.\ Sloan Foundation, the Participating
Institutions, the National Science Foundation, the U.S.\ Department of
Energy, the National Aeronautics and Space Administration, the
Japanese Monbukagakusho, the Max Planck Society, and the Higher
Education Funding Council for England.

\end{acknowledgements}

\appendix

\section{ASSESSING THE INTEGRATED \hi\ FLUX COMPLETENESS OF THE \alf\ DATA}\label{completeness}

Although the \alf\ catalog and its LDE subset are noise-limited
datasets, it is possible to define an integrated flux limit, $\Flim$,
above which the confirmed \hi\ sources are not subject on average to
the same bias against broad linewidths.

With this aim, we have used the adaptation of the \citet{Rau01}
completeness test to a \hi-selected galaxy sample by \citet{Zwa04},
which we briefly recap here. \citeauthor{Rau01}'s method, which is not
affected by the presence of clustering or by subsampling in redshift
bins, relies on the calculation for each galaxy $i$ of the quantity
\begin{equation}\label{Rauzy_z}
\hat{\zeta}_i=\frac{r_i}{n_i+1}\;, 
\end{equation} 
which provides an unbiased estimate of the random variable $\zeta_i$ that
compares the amount of galaxies with more and less neutral hydrogen
than every galaxy in the sample under the assumption that the shape of
the \hi\ mass function is universal (i.e., invariant in time and
position). In equation~\ref{Rauzy_z}, $r_i$ is the number of galaxies
with $\Mhi\geq M_{\rm{\scriptstyle{H}\,\scriptscriptstyle I},i}$ and
$Z\leq Z_i$, $n_i$ is the number of galaxies for which $\Mhi\geq
M_{\rm{\scriptstyle{H}\,\scriptscriptstyle I},lim}(Z_i)$ and $Z\leq
Z_i$, while $Z\equiv\log\Fhi -\log\Mhi$ is a distance measure, and
$M_{\rm{\scriptstyle{H}\,\scriptscriptstyle I},lim}(Z_i)$ is the
limiting \hi\ mass at the distance corresponding to $Z_i$.

Taking into account that $\zeta_i$ is uniformly distributed between 0 and
1 with expectation and variance $E_i=1/2$ and
$V_i=(n_i-1)/[12(n_i+1)]$, respectively, the principle of the test is
to evaluate the variation with decreasing $\Flim$ of the statistic
\begin{equation}
T_{\rm C}=\frac{\sum_{i=1}^{N_{\rm gal}}(\hat{\zeta}_i-1/2)}{\left(\sum_{i=1}^{N_{\rm gal}}V_i\right)^{1/2}}\;,
\end{equation}
which follows a Gaussian distribution of zero mean and unit variance
under the null hypothesis, H0, that the sample is complete up to a
given integrated flux and drops systematically to values below zero
when H0 is not fulfilled. The completeness limit can be therefore set
by imposing that $T_{\rm C}$ exceeds a given negative value, being
$T_{\rm C}<-2$ and $T_{\rm C}<-3$, which correspond to a 97.7 and 99.4
confidence levels of rejection of H0, respectively, the standard
decision rules.

\begin{figure}
\begin{center}
\epsscale{0.8}
\plotone{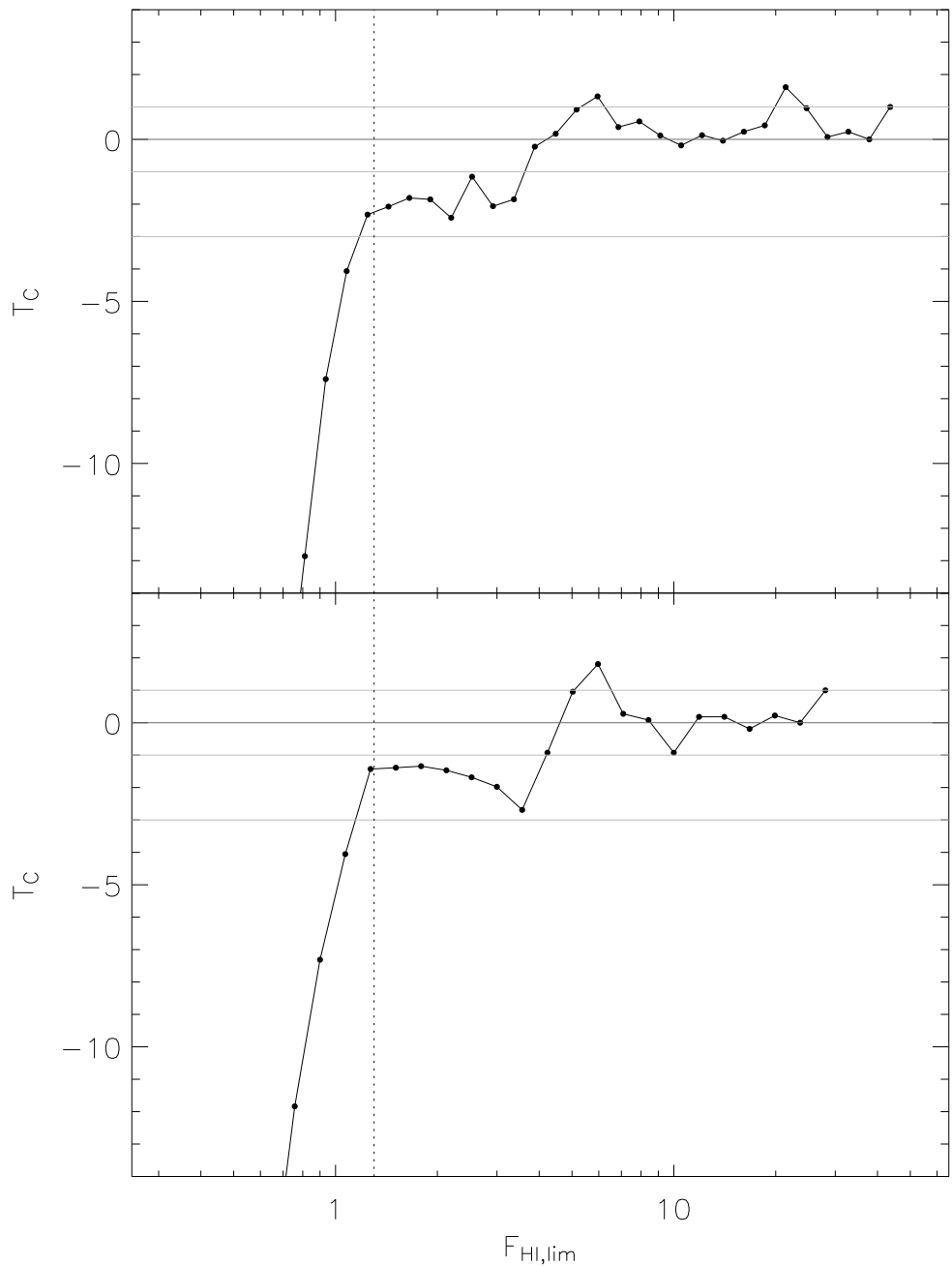}
\end{center}
\caption{\citeauthor{Rau01}'s test for completeness applied to data
  from the \alf\ survey (top) and from its LDE subset (bottom). We
  only consider code 1 sources with redshifts between 2000 and 15,000
  \kms. In both cases, the \hi\ flux completeness limit, $\Flim$,
  (determined by $T_{\rm C}=-3$) falls somewhat below $1.3$ Jy~\kms\
  (vertical dotted line).}\label{fig_rauzy}
\end{figure}

Figure~\ref{fig_rauzy} shows the results obtained when the test is
applied to subsamples of both the \alf\ data without any density
restriction (top panel) and the LDE dataset (bottom panel) truncated to
decreasing values of $\Flim$. Note that for both datasets we are
considering only code 1 sources within the bandpass limits 2000 and
15,000 \kms\ (see Section~\ref{data_selection}), a subsampling in
redshift that should not affect the outcome. It is clear from this
figure that if we choose a $-3\sigma$ criterion to reject the
completeness hypothesis, which corresponds to the level from which the
$T_{\rm C}$ statistic initiates a systematic, sharp decline, the
completeness limit of the two samples can be safely set at $1.3$
Jy~\kms.

A more direct, but less accurate, method to infer whether a survey is
complete or not to a given flux limit is to compute the average
$V/V_{\rm max}$ of the data, which should be equal to $1/2$
\emph{provided that the effective search volume of each galaxy, or
  equivalently, its detection probability, is accurately
  established}. The $\langle V/V_{\rm max}\rangle $ test presupposes
that the galaxies are on average homogeneously distributed and is
therefore sensitive to selection and large-scale structure
effects. Therefore, the correct application of this second technique
to the above \hi\ datasets requires that in the calculation of the
individual search volumes we account for the true density of targets
as a function of redshift, as well as for the artificial annular
underdensities that arise from man-made radio-frequency interferences.

Accordingly, we have modified $V$ and $V_{\rm max}$ in order to
include weighting by the average density interior to the corresponding
radial distances normalized to the average density within the surveyed
volume. The weights have been calculated from a volume-limited
subsample of the spectroscopic SDSS DR7 dataset, which is complete to
a Petrosian $r$-band magnitude of 17.77, selected to include only
galaxies obeying \citet{Mal09}'s disk criterion, which is very
effective identifying objects structurally similar to \hi\ emitters
(see Paper I). In addition, we have corrected for RFI by using the
same average relative weight as a function of observed heliocentric
velocity depicted in Figure~6 by \citet{Mar10}. 

\begin{figure}
\begin{center}
\epsscale{0.8}
\plotone{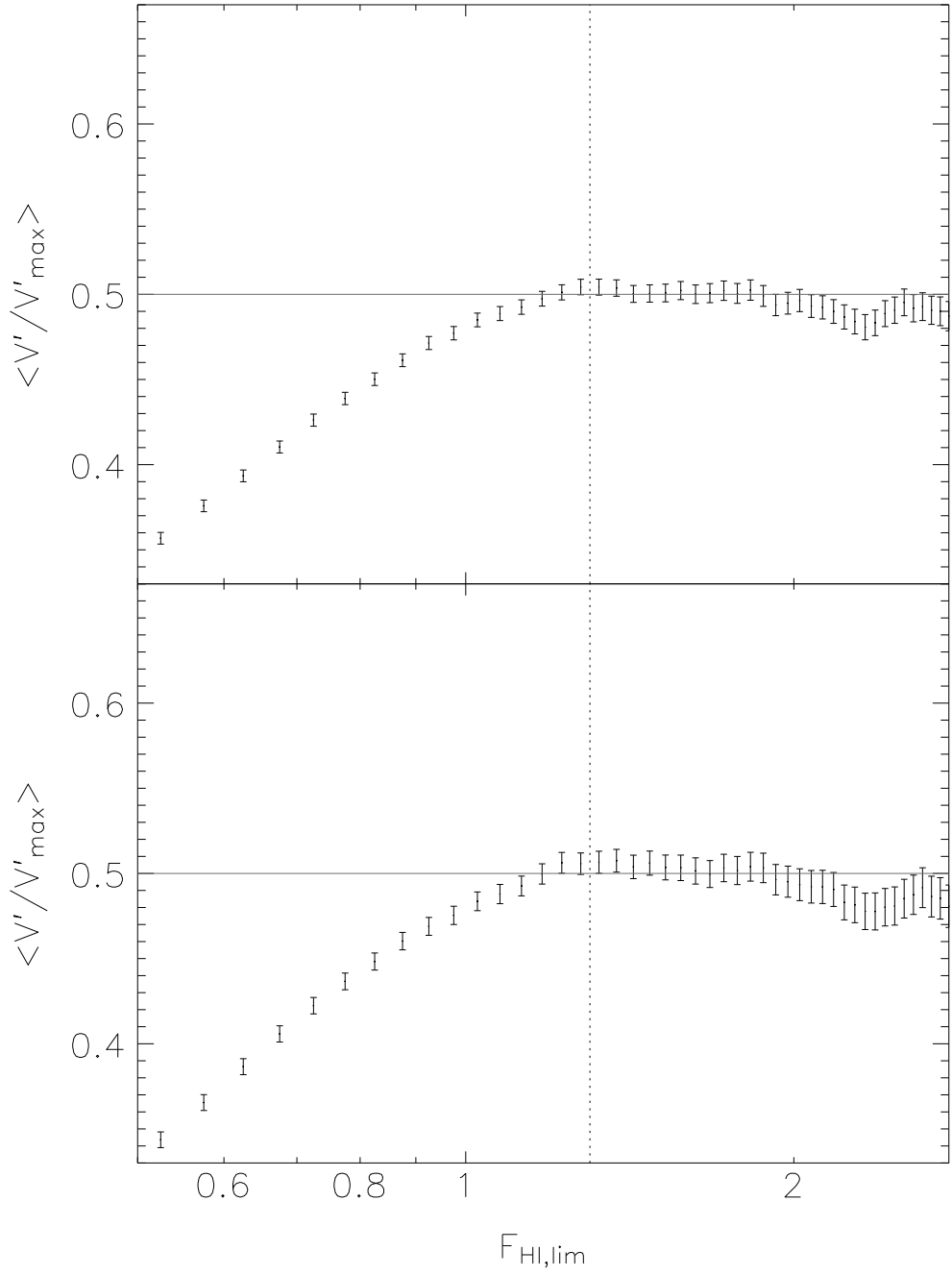}
\end{center}
\caption{Same as in Figure~\ref{fig_rauzy}, but for the $\langle
  V^\prime/V^\prime_{\rm max}\rangle$ completeness
  test.}\label{fig_v_over_vmax}
\end{figure}

In Figure~\ref{fig_v_over_vmax}, we depict the expectation value and
dispersion of the average ratio of weighted volumes, $\langle
V^\prime/V^\prime_{\rm max}\rangle$, as a function of the integrated
\hi\ flux. As we have done previously for \citeauthor{Rau01}'s test,
we show results for the \alf\ data (top panel) and its LDE subset
(bottom panel; in this case, the radial run of the density weighting
is estimated using \citeauthor{Mal09}'s disks in low density
environments). Galaxies not belonging to the redshift range
2000--$15,000$ \kms\ or classified as code 2 detections have been
discarded. In very good agreement with the $T_{\rm C}$ method, we find
that above $\Flim\approx 1.3$ Jy~\kms\ the values of this statistic
remain, for the two datasets, practically constant around 0.50. Note
that the observed rough invariance of $\langle V^\prime/V^\prime_{\rm
  max}\rangle$ above this integrated-flux limit supports the assertion
that the two samples are statistically complete, while the fact that
the observed value of the statistic is so close to its expectation
indicates that the effects of large-scale structure and RFI have been
correctly averaged out with the adopted weighting strategy and,
therefore, that we are dealing with homogeneous datasets.




\begin{deluxetable}{lrrrrr|rr}
\tabletypesize{\scriptsize}
\tablecolumns{8}
\tablewidth{0pt}
\tablecaption{Principal Component Analysis of \hi\ and Optical Properties for Weighted Data and Disattenuated Correlations\label{tab_pca_att}}
\tablehead{
&
\colhead{$\log \Mhi$} &
\colhead{$\log \W$} &
\colhead{$\log \rdiam$} & 
\colhead{$M_r$} &
\colhead{$(g-r)$}&
\colhead{$\rici$}&
\colhead{$\rba$}
}
\startdata

$\log \Mhi $  &     1.00 &     0.65 &     0.80 &  $-$0.74 &     0.58 &  $-$0.18 &     0.11  \\
$\log \W$     &          &     1.00 &     0.71 &  $-$0.76 &     0.88 &  $-$0.22 &     0.08  \\
$ \log \rdiam$&          &          &     1.00 &  $-$0.96 &     0.85 &  $-$0.09 &     0.02  \\
$M_r$         &          &          &          &     1.00 &  $-$0.87 &     0.25 &  $-$0.00  \\
$(g-r)$       &          &          &          &          &     1.00 &  $-$0.19 &     0.20  \\
$\rici$       &          &          &          &          &          &     1.00 &     0.10  \\
$(b/a)_r$     &          &          &          &          &          &          &     1.00 \\

Mean ($N_{\rm gal}=1624$) &8.82     & 2.22   & 0.42    &$-$17.43     & 0.33     & 0.44   &   0.50 \\

Standard dev.\ & 0.36  &     0.18   &    0.23    &   1.48 &      0.12  &     0.05   &    0.16\\

Observational error & 0.02  &    0.03   &   0.05  &    0.10    &  0.10    &  0.01   &   0.07\\

$\sqrt{R_{XX}}$ &     1.00 &     0.99 &     0.98 &     1.00 &     0.79 &     0.98 &     0.92  \\

\hline
  \multicolumn{6}{c}{Eigenvectors}& \multicolumn{2}{c}{Eigenvalues (\%)}\\
\hline

$\univ_1$ &     0.40 &     0.43 &     0.47 &  $-$0.47 &     0.46 &   $\lambda_1=$  4.17 &   (83.31)  \\
$\univ_2$ &     0.73 &  $-$0.45 &     0.21 &  $-$0.03 &  $-$0.46 &   $\lambda_2=$  0.49 &   (93.02)  \\
$\univ_3$ &     0.46 &     0.63 &  $-$0.43 &     0.43 &  $-$0.12 &   $\lambda_3=$  0.30 &   (99.07)  \\
$\univ_4$ &  $-$0.08 &     0.21 &  $-$0.41 &  $-$0.75 &  $-$0.47 &   $\lambda_4=$  0.05 &   (100)  \\
$\univ_5$ &  $-$0.29 &     0.40 &     0.62 &     0.17 &  $-$0.58 &   $\lambda_5=$  0.00 &   (100)  \\

\hline
rms residuals: &&&&&& \\
\ \ \ Principal Axis &     0.22 &     0.09 &     0.09 &     0.51 &     0.09  \\
\ \ \ Principal Plane&     0.11 &     0.09 &     0.07 &     0.50 &     0.06  \\
\ \ \ Princ.\ 3-Plane&     0.06 &     0.05 &     0.05 &     0.42 &     0.06  \\
\ \ \ Princ.\ 4-Plane&     0.05 &     0.04 &     0.07 &     0.12 &     0.04  \\
\ \ \ Princ.\ 5-Plane&     0.00 &     0.00 &     0.00 &     0.00 &     0.00  \\

\enddata 
\tablecomments{PCA is carried on the first 5 properties only: $\log (\Mhi\,[\msun])$, 
$\log (\W\,[\mbox{\kms}])$, $\log (\rdiam\,[\mbox{kpc}])$, $M_r\,[\mbox{mag}]$, and $(g-r)\,[\mbox{mag}]$.}
\end{deluxetable}

\begin{deluxetable}{lrrrrr|rr}
\tabletypesize{\scriptsize}
\tablecolumns{8}
\tablewidth{0pt}
\tablecaption{Principal Component Analysis of \hi\ and Optical Properties for Weighted Data\label{tab_pca}}
\tablehead{
&
\colhead{$\log \Mhi$} &
\colhead{$\log \W$} &
\colhead{$\log \rdiam$} & 
\colhead{$M_r$} &
\colhead{$(g-r)$}&
\colhead{$\rici$}&
\colhead{$\rba$}
}
\startdata

 $\log \Mhi$     &     1.00 &     0.64 &     0.78 &  $-$0.73 &     0.46 &  $-$0.17 &     0.10  \\
 $\log \W$       &          &     1.00 &     0.69 &  $-$0.75 &     0.69 &  $-$0.21 &     0.07  \\
 $\log \rdiam$   &          &          &     1.00 &  $-$0.94 &     0.66 &  $-$0.09 &     0.02  \\
 $M_r$           &          &          &          &     1.00 &  $-$0.69 &     0.24 &  $-$0.00  \\
 $(g-r)$         &          &          &          &          &     1.00 &  $-$0.15 &     0.14  \\
 $\rici$         &          &          &         &           &         &     1.00  &     0.09\\
 $\rba$          &          &          &         &           &         &           &     1.00\\

Mean ($N_{\rm gal}=1624$) &8.82     & 2.22   & 0.42    &$-$17.43     & 0.33     & 0.44   &   0.50 \\

Standard dev.\ & 0.36  &     0.18   &    0.23    &   1.48 &      0.12  &     0.05   &    0.16\\



\hline
  \multicolumn{6}{c}{Eigenvectors}& \multicolumn{2}{c}{Eigenvalues (\%)}\\
\hline

$\univ_1$ &     0.42 &     0.44 &     0.48 &    $-$0.48 &       0.41 &   $\lambda_1=$  3.83 &    (76.57)  \\
$\univ_2$ &     0.61 &  $-$0.25 &     0.21 &    $-$0.09 &    $-$0.72 &   $\lambda_2=$  0.57 &    (88.06)  \\
$\univ_3$ &  $-$0.25 &  $-$0.78 &     0.43 &    $-$0.31 &       0.23 &   $\lambda_3=$  0.33 &    (94.60)  \\
$\univ_4$ &  $-$0.61 &     0.34 &     0.20 &    $-$0.45 &    $-$0.52 &   $\lambda_4=$  0.22 &    (98.98)  \\
$\univ_5$ &  $-$0.14 &     0.15 &     0.71 &       0.67 &    $-$0.05 &   $\lambda_5=$  0.05 &   (100)  \\

\hline
rms residuals: &&&&&& \\
\ \ \ Principal Axis  &     0.20 &     0.09 &     0.08 &     0.48 &     0.10&&  \\
\ \ \ Principal Plane &     0.12 &     0.09 &     0.07 &     0.47 &     0.04&&  \\
\ \ \ Princ.\ 3-Plane &     0.10 &     0.03 &     0.04 &     0.39 &     0.04&&  \\
\ \ \ Princ.\ 4-Plane &     0.01 &     0.01 &     0.04 &     0.23 &     0.00&&  \\
\ \ \ Princ.\ 5-Plane &     0.00 &     0.00 &     0.00 &    0.00  &    0.00&&\\

\enddata 
\tablecomments{PCA is carried on the first 5 properties only: $\log (\Mhi\,[\msun])$, 
$\log (\W\,[\mbox{\kms}])$, $\log (\rdiam\,[\mbox{kpc}])$, $M_r\,[\mbox{mag}]$, and $(g-r)\,[\mbox{mag}]$.}
\end{deluxetable}


\begin{deluxetable}{l|l|r|rrrrrrrrr|c}
\rotate
\tablecolumns{13}
\tablewidth{0pt}
\tablecaption{Coefficients of $\Mhi$ Predictions from Single and Multiple Linear Regression Models\label{tab_pred}}
\tablehead{
Weighting &\multicolumn{1}{c|}{$X_1$}  &\multicolumn{1}{c|}{$X_2$}  &\multicolumn{3}{c}{$a_0$}  & \multicolumn{3}{c}{$a_1$} & \multicolumn{3}{c|}{$a_2$} & Residual
}
\startdata
\multirow{5}{*}{$1/V^\prime_{\rm max}$}& $\log \rdiam$ & & 8.72 &$\pm 0.06$ & $\pm 0.06$  & 1.25 &$\pm 0.06$& $\pm 0.07$& & & & 0.23 \\

&  $M_r$  & &6.44 & $\pm 0.20$ & $\pm 0.21$ & $-$0.18 &$\pm 0.01$& $\pm 0.01$  & && & 0.25 \\

&  $\log \W$ & &6.54 & $\pm 0.27$ & $\pm 0.20$ & 1.30& $\pm 0.11$ & $\pm 0.09$ &  & &&  0.28\\

&  $(g-r)$   & & 8.84 & $\pm 0.11 $ & $\pm 0.12$ & 1.81 &$\pm 0.29$ & $\pm 0.40$ &  &&& 0.33 \\

&  $\log \rdiam$ & $M_r$ &7.26 &  $\pm 0.12$ & $\pm 0.04$ & 0.66 &  $\pm 0.03$& $\pm 0.01$  & $-$0.10&  $\pm 0.006$ & $\pm 0.002$  & 0.22\\

\hline
\multirow{5}{*}{None}& $\log \rdiam$& & 8.85 & $\pm 0.04$ & $\pm 0.03$  & 1.37 &$\pm 0.04$ & $\pm 0.03$ &&& & 0.21 \\

&  $M_r$ & &  6.44 & $\pm 0.09$ & $\pm 0.08$ & $-$0.20 &$\pm 0.004$& $\pm   0.002$  & && & 0.23 \\

&  $\log \W$ & & 7.17  & $\pm 0.14$ & $\pm 0.16$ & 1.21&$\pm 0.05$& $\pm   0.06$ &  & && 0.28 \\

&  $(g-r)$ & &9.61 & $\pm 0.04$ & $\pm 0.04$ & 1.10& $\pm 0.08$ & $\pm   0.07$ &  & && 0.32 \\

&  $\log \rdiam$ & $M_r$ & 6.89 &$\pm 0.05$ &$\pm 0.02$ &0.61 &$\pm  0.01$& $\pm 0.005$& $-$0.10 & $\pm 0.002$ & $\pm 0.001$&0.23\\

\enddata
\end{deluxetable}

%
%
\begin{deluxetable}{l|l|rr|rr|rr|rr}
\tabletypesize{\scriptsize}
\rotate
\tablecolumns{10}
\tablewidth{0pt}
\tablecaption{Coefficients of Orthogonal Fits between Pairs of Variables\label{tab_sca}} 
\tablehead{& & \multicolumn{8}{|c}{$X$}\\
\cline{3-10}
Weighting & \multicolumn{1}{c}{$Y$}& \multicolumn{2}{|c}{$\log \rdiam$} & \multicolumn{2}{|c}{$M_r$}  & 
\multicolumn{2}{|c}{$\log \W$} & \multicolumn{2}{|c}{$(g-r)$}\\
\cline{3-10}
 &  &\multicolumn{1}{|c}{$a_0$} & \multicolumn{1}{c}{$a_1$} &\multicolumn{1}{|c}{$a_0$} & \multicolumn{1}{c}{$a_1$} &\multicolumn{1}{|c}{$a_0$} &\multicolumn{1}{c}{$a_1$}&\multicolumn{1}{|c}{$a_0$} & \multicolumn{1}{c}{$a_1$}
}
\startdata
\multirow{4}{*}{$1/V^\prime_{\rm max}$}&$\log \Mhi$ & 8.55$\pm 0.05$  & 1.55$\pm 0.06$ & 5.36$\pm 0.19$ &  $-$0.24$\pm 0.010$ & 5.01$\pm 0.30$ & 1.99$\pm 0.13$ & 8.45$\pm 0.12$ & 2.99$\pm 0.29$  \\

&$\log \rdiam$ & & &  $-$2.05$\pm 0.09$ &$-$0.16$\pm 0.004$  & $-$2.28$\pm 0.21$ & 1.29$\pm 0.14$& $-$0.06$\pm 0.06$ &1.93$\pm 0.15$\\

& $M_r$ & & & & & 1.46$\pm 1.14$ & $-$8.15$\pm 0.48$ &$-$12.8$\pm 0.40$ &$-$12.2$\pm 0.98$\\

& $\log \W$ &  & & &  & & & 1.73$\pm 0.06$&1.50$\pm 0.13$\\
\hline
\multirow{4}{*}{None}&$\log \Mhi$& 8.58$\pm 0.03$  & 1.66$\pm 0.03$ & 5.24$\pm 0.08$ & $-$0.26$\pm 0.004$  & 5.30$\pm 0.11$ & 1.98$\pm 0.04$ & 9.06$\pm 0.04$ & 2.28$\pm 0.07$  \\

&$\log \rdiam$ & & & $-$2.02$\pm 0.03$ & $-$0.16$\pm 0.002$ & $-$1.98$\pm 0.08$ & 1.19$\pm 0.03$ & 0.29$\pm 0.02$& 1.38$\pm 0.03$\\

&$M_r$ & & & & & $-$0.24$\pm 0.42$  &$-$7.57$\pm 0.16$  & $-$14.6$\pm 0.13$ &$-$8.74$\pm0.26$\\

& $\log \W$ &  & & &  & & &1.90$\pm 0.01$& 1.15$\pm 0.04$\\
\enddata

\end{deluxetable}


\begin{deluxetable}{l|ccccc}
\tablecolumns{6}
\tablewidth{0pt}
\tablecaption{Central Slopes of Scaling Laws between Fundamental Galaxian Properties Reported by Different Authors\label{tab_sca_fun}}
\tablehead{\multicolumn{1}{c|}{Reference}&\multicolumn{5}{c}{Scaling law}\\
\cline{2-6}
& \multicolumn{1}{c}{$\Mhi\sim R^\alpha$}
& \multicolumn{1}{c}{$\Mhi\sim L^\beta$}& \multicolumn{1}{c}{$L\sim V^\gamma$}& \multicolumn{1}{c}{$R\sim L^\delta$}& \multicolumn{1}{c}{$R\sim V^\epsilon$}
}
\startdata
\citet{HG84}&1.8&0.66&2.6& \nodata  & \nodata  \\
\citet{SH96}&2.0&0.74&3.7& 0.37& 1.4\\
\citet{Cou07}& \nodata  &\nodata  &3.4&0.32&1.1\\
This work&1.6&0.60&3.3&0.40 &1.3\\ 
\enddata


\end{deluxetable}

\clearpage

\end{document}